\documentclass[showkeys,prc,floatfix,nofootinbib, showpacs]{revtex4}
\usepackage{amssymb}
\usepackage{graphicx}
\usepackage{bm}
\usepackage{latexsym}
\usepackage[english]{babel}
\usepackage{color}
\begin{document}
\newcommand{\ti}[1]{\mbox{\tiny{#1}}}
\newcommand{\im}{\mathop{\mathrm{Im}}}
\def\be{\begin{equation}}
\def\ee{\end{equation}}
\def\bea{\begin{eqnarray}}
\def\eea{\end{eqnarray}}

\title{ON THE EQUIVALENCE OF APPROXIMATE STATIONARY AXIALLY SYMMETRIC 
SOLUTIONS OF EINSTEIN FIELD EQUATIONS}

\author{Kuantay Boshkayev$^{1,2}$,  Hernando Quevedo$^{2,3}$, Saken Toktarbay$^1$, Bakytzhan Zhami$^1$ and Medeu Abishev $^{1, 4}$ }
\email{kuantay@mail.ru, quevedo@nucleares.unam.mx, saken.yan@yandex.com, zhami.bakytzhan@gmail.com, abishevme@mail.ru}
\affiliation{
$^1$Department of Theoretical and Nuclear Physics, Al-Farabi Kazakh National University,  Almaty 050040, Kazakhstan \\
$^2$Dipartimento di Fisica and ICRA, Universit\`a di Roma ``La Sapienza", I-00185 Roma, Italy \\
$^3$Instituto de Ciencias Nucleares, Universidad Nacional Aut\'onoma de M\'exico, AP 70543, M\'exico, DF 04510, Mexico \\
$^4$ Institute of Gravitation and Cosmology, Peoples Friendship University of Russia (PFUR), Moscow 117198, Russia }

\date{\today}

\begin{abstract}
We study stationary axially symmetric solutions of the Einstein vacuum field equations that can be used to describe the gravitational field of astrophysical compact objects in the limiting case of slow rotation and slight deformation. We derive explicitly the exterior Sedrakyan-Chubaryan approximate solution, and express it 
in analytical form, which makes it practical in the context of astrophysical applications. In the limiting case of vanishing angular momentum, the solution reduces 
 to the well-known Schwarzschild solution in vacuum. We demonstrate that the new solution is equivalent to the exterior Hartle-Thorne solution. We establish the mathematical equivalence between the Sedrakyan-Chubaryan, Fock-Abdildin and Hartle-Thorne formalisms.

\end{abstract} 
\pacs{xx.xx.xx;xx.xx.xx}

\maketitle

\section{Introduction}
\label{sec:int}

To study the gravitational field of slowly uniformly rotating and slightly deformed relativistic objects, Hartle developed in his original work \cite{hartle1967}  a method in the slow rotation approximation, extending the well-known exterior and interior Schwarzschild solutions. The method allows one to investigate the physical properties of rotating stellar objects in hydrostatic equilibrium. It was first applied to real astrophysical objects by Hartle and Thorne \cite{hartlethorne1968}, employing the Harrison-Wheeler,  Tsuruta-Cameron and the Harrison-Wakano-Wheeler equations of state. Soon after, the method has become known as the Hartle-Thorne 
approach, and there appeared a new series of research papers extending, modifying and improving the original approach by including higher order multipole moments and corrections in the angular momentum, etc. \cite{stergioulas2003, yagi2014}. Furthermore, the Hartle formalism was tested, compared and contrasted with numerical computations in full general relativity \cite{berti2004, berti2005}. As a result,  it was shown that the Hartle formalism can be safely used to study stellar objects with intermediate rotation periods. Only for higher angular velocities, close to the mass-shedding limit, it shows noticeable discrepancies from the full general relativistic simulations \cite{stergioulas2003, berti2004, berti2005}.

Similar approaches were developed by Bradley et. al. in \cite{bradley2000, fodor2002, bradley2007, bradley2009}, where the slow rotation approximation is used in order to construct interior and exterior solution to the Einstein field equations. Unlike Hartle, Bradley et. al solved the six independent Einstein equations without involving the integral of the equation of hydrostatic equilibrium for uniformly rotating configurations. Moreover, the Darmois-Israel procedure was applied to match the interior and exterior solutions. In some particular cases,  Bradley et. al. \cite{bradley2000, fodor2002, bradley2007, bradley2009} included the electric charge by solving the Einstein-Maxwell equations.

In addition, Konno et. al \cite{konno1999} generalized Hartle's approach in the static case to include the deformation of relativistic stars due to the presence of magnetic fields. Afterwards,  Konno and coworkers \cite{konno2000} calculated the ellipticity of the deformed stars due to the presence of both magnetic field and rotation, extending their previous results. This method has become popular and found its astrophysical application in the physics of all types of magnetic stars 
\cite{konno2001, ioka, colaiuda2008, mallick2014, folomeev}.

On the other hand, independently of Hartle, Sedrakyan and Chubaryan \cite{sedrakyan1968} formulated their own distinctive approach for calculating the 
exterior gravitational
structure of equilibrium rigidly rotating superdense stars in the small angular velocity approximation, though it is not well-known in the scientific community. 
The corresponding interior solution, together with the matching procedure, was obtained in their subsequent paper \cite{sedrakyan21968}. 
The manner of solving the Einstein equations was markedly different from the Hartle's approach. Further applications of the Sedrakyan-Chubaryan solution to white dwarfs and neutron stars were considered in a number of papers e.g. \cite{arutyunyan1971, arutyunyan21971, arutyunyan1973}. Numerical results obtained by Arutyunyan et. al \cite{arutyunyan21971} were in agreement with the ones computed by Hartle and Thorne \cite{hartlethorne1968}, implying that there was no contradiction between these two solutions.

Besides, the exterior Sedrakyan-Chubaryan solution was written in an analytic form 
\cite{sedrakyan1968, sedrakyan21968, arutyunyan1971, arutyunyan21971, arutyunyan1973}, and it required the additional integration of one of the metric functions, under a careful consideration of the boundary conditions. Maybe this was one of the main causes that the Sedrakyan-Chubaryan solution is still less known in the scientific community. Indeed, this fact does not allow one to compare and contrast it with the exterior Hartle-Thorne solution straightforwardly.  The main goal of the present work is to derive explicitly the exterior Sedrakyan-Chubaryan solution and to establish
its relationship with the Hartle-Thorne solution. In fact, we will show that they are related by means of a coordinate transformation, whose non-trivial part includes only the radial coordinate, and a redefinition of the parameters entering the solution. 

This paper is organized as follows. In Sec. \ref{sec:ht}, we present the explicit form of the exterior Hartle-Thorne metric. In Sec. \ref{sec:sc}, we use a particular 
line element to derive explicitly all the components of the stationary Sedrakyan-Chubaryan metric up to the second order in the angular velocity. In Sec. \ref{sec:tra},
we find the transformation that establishes the equivalence of the two metrics under consideration. Finally, in Sec. \ref{sec:con}, we review our results. 
We will follow the notation of \cite{sedrakyan1968}, and use the geometric units with $G=c=1$ throughout the paper. 



\section{The Hartle-Thorne approximate solution}
\label{sec:ht}

The exterior Hartle-Thorne  metric describes the gravitational field of a slowly rotating slightly deformed source in vacuum. 
In geometric units, the metric is given by \cite{hartle1967}

\begin{eqnarray}\label{HT}
ds^2&=&-\left(1-\frac{2{ M }}{r}\right)\left[1+2k_1P_2(\cos\theta)+2\left(1-\frac{2{M}}{r}\right)^{-1}\frac{J^{2}}{r^{4}}(2\cos^2\theta-1)\right]dt^2 \nonumber \\ 
&&+\left(1-\frac{2{M}}{r}\right)^{-1}\left[1-2k_2P_2(\cos\theta)-2\left(1-\frac{2{M}}{r}\right)^{-1}\frac{J^{2}}{r^4}\right]dr^2\\
&&+r^2[1-2k_3P_2(\cos\theta)](d\theta^2+\sin^2\theta d\phi^2)-\frac{4J}{r}\sin^2\theta dt d\phi \nonumber\,
\end{eqnarray}
where

\begin{eqnarray}\label{HTk1}
k_1&=&\frac{J^{2}}{{M}r^3}\left(1+\frac{{M}}{r}\right)+\frac{5}{8}\frac{Q-J^{2}/{M}}{{M}^3}Q_2^2\left(\frac{r}{{M}}-1\right), \quad k_2=k_1-\frac{6J^{2}}{r^4}, \\ \nonumber
k_3&=&k_1+\frac{J^{2}}{r^4}+\frac{5}{4}\frac{Q-J^{2}/{M}}{{M}^2r}\left(1-\frac{2{M}}{r}\right)^{-1/2}Q_2^1\left(\frac{r}{M}-1\right), \nonumber\\
P_{2}(x)&=&\frac{1}{2}(3x^{2}-1),\nonumber \\
Q_{2}^{1}(x)&=&(x^{2}-1)^{1/2}\left[\frac{3x}{2}\ln\frac{x+1}{x-1}-\frac{3x^{2}-2}{x^{2}-1}\right],\nonumber\\ Q_{2}^{2}(x)&=&(x^{2}-1)\left[\frac{3}{2}\ln\frac{x+1}{x-1}-\frac{3x^{3}-5x}{(x^{2}-1)^2}\right].\nonumber
\end{eqnarray}
Here $P_{2}(x)$ is Legendre polynomials of the first kind, $Q_l^m$ are the associated Legendre polynomials of the second kind and the constants ${M}$, ${J}$ and ${Q}$ are the total mass, angular momentum and  quadrupole moment of the rotating source, respectively.

Unlike other solutions of the Einstein equations, the Hartle-Thorne solution has an internal counterpart, which makes it more practical with respect to the exact solutions. All the internal functions are interrelated with the external ones. Thus, the total mass, angular momentum and quadrupole moment of a rotating star are determined through the constants obtained by means of the numerical integration of both  interior and exterior solutions, by applying the matching conditions
on the surface of the star.

\section{The Sedrakyan-Chubaryan solution}
\label{sec:sc}

In this section, we derive the approximate Sedrakyan-Chubaryan solution \cite{sedrakyan1968} in detail. We will limit ourselves to the 
exterior solution for which we derive all the metric functions. 


Following the procedure presented in \cite{sedrakyan1968}, we consider the  line element for axially symmetric rotating stars in the form
\be
ds^2 = \left( \omega^2 e^\mu \sin^2 \theta - e^\nu \right)dt^2+e^\lambda dr^2+ e^\mu \left( d\theta^2 + \sin^2 \theta d\phi^2 \right)+ 2 \ \omega \ e^\mu \sin^2 \theta \ d \phi dt  ,
\label{lel}
\ee
where  $\lambda=\lambda(r,\theta)$, \ $\mu =\mu (r,\theta)$, \ $\omega=\omega(r,\theta)$ and $\nu=\nu(r,\theta)$ are functions of the radial $r$ and 
angular $\theta$ coordinates.  Note that, $\omega$ is proportional to the odd powers of the angular velocity $\Omega$, whereas the remaining functions are proportional to the even powers of $\Omega$. We will consider here an approximation up to the second order in $\Omega$. We now demand that the above
metric satisfies vacuum Einstein's equations in the form 
\be
G_{\beta}^{\alpha}=R_{\beta}^{\alpha} - \frac{1}{2} R \delta_{\beta}^{\alpha} =  0 \ . 
\label{eins}
\ee

In the limiting case of a static star, the angular velocity  $\Omega=0$ and the function $\omega=0$; then, $\lambda, \nu$ and $\mu$ are  functions of the radial coordinate $r$ only. Obviously, for this special case we automatically obtain the exterior Schwarzschild solution
\bea 
e^\nu&=&e^{\nu_{0}}= \left( 1-\frac{2 m}{r} \right), \\
e^\lambda&=& e^{\lambda_{0}}= \left( 1-\frac{2 m}{r} \right)^{-1},\\
e^\mu&=&e^{\mu_{0}}= r^2 ,
\eea
where $m$ is the static mass.

We now consider the line element (\ref{lel}) for a slowly rotating relativistic star. In this case, we can expand the functions $\lambda$, $\mu$, $\nu$ and $\omega$ in powers of the angular velocity of the star $\Omega$, assuming that $\Omega$ is small. As a parameter for Taylor expanding the metric tensor components, it is convenient to introduce the dimensionless quantity $\beta=\Omega^2/ 8 \pi \rho_c$, where $\rho_c$ is the central density of the configuration. Thus, we define the metric functions as
\bea 
e^{\nu \left(r, \theta \right)}&=& e^{\nu_{0}}\left[1+ \beta\Phi \left(r, \theta \right)\right], \\
e^{\lambda \left(r, \theta \right)}&=& e^{\lambda_{0}}\left[ 1- \beta f \left(r, \theta \right)\right], \\
e^{\mu \left(r, \theta \right)}&=& e^{ \mu _{0}}\left[1+ \beta U \left(r, \theta \right) \right], \\
\omega \left(r, \theta \right) &=& \sqrt{\beta} q \left(r \right) \ ,
\eea
where the functions \ $\mu_{0}$,\ $\nu_{0}$ and $\lambda_{0}$ represent the Schwarzschild solution, and $U$, $\Phi$, and $f$ are unknown functions.
To find the independent differential equations from the Einstein field equations, we make use of the following combinations
\be 
G_{1}^{1}-G_{0}^{0}=0, \ \ \ G_{2}^{2}+G_{3}^{3}=0, \ \ \ G_{2}^{1}=0, \ \ \ G_{0}^{3}=0.
\label{con_eqn}
\ee

In order to solve each component of Eq.(\ref{con_eqn}), all the metric functions are expanded in spherical harmonics. In turn, this procedure allows one to separate variables. Retaining only the terms responsible for the quadrupolar deformation, we have
\bea 
\Phi\left(r,\theta \right)&=&\sum_{l=0}^\infty \Phi_l \left(r \right) P_l \left( \cos \theta \right)\approx \Phi_{0} \left(r \right)P_{0} \left(\cos \theta \right)+\Phi_{2} \left(r \right) P_{2} \left(\cos \theta \right),  
\label{spher_Phi}\\
f\left(r,\theta \right)&=&\sum_{l=0}^\infty f_l \left(r \right) P_l \left( \cos \theta \right)\approx f_{0} \left(r \right)P_{0} \left(\cos \theta \right)+f_{2} \left(r \right) P_{2} \left(\cos \theta \right),
\label{spher_f}\\
U\left(r,\theta \right)&=&\sum_{l=0}^\infty U_l \left(r \right) P_l \left( \cos \theta \right)\approx U_{0} \left(r \right)P_{0} \left(\cos \theta \right)+U_{2} \left(r \right) P_{2} \left(\cos \theta \right),
\label{spher_U}
\eea
where $P_{0} (\cos \theta)$ and $P_{2} (\cos \theta)$  are the Legendre polynomials of the first kind 
\bea
P_{0} (\cos \theta)=1, \ \ P_{2} (\cos \theta)= -\frac{1}{2} \left(1-3 \cos^2 \theta \right).
\label{P02}
\eea
Note that because the axis of symmetry is oriented along the rotation axis, the expansion in spherical harmonics contains only even values of $l$. Moreover, in the slow rotation approximation $l$ accepts only two values: $l=0$ and $l=2$. 
The components of the Einstein tensor $G^0_3$ or $G^3_0$  yield a differential equation which is proportional to $\Omega$,
\be 
q_{,rr}+ \frac{4 \ q_{,r}}{r}=0,
\ee 
where $q_{,r}= \frac{\partial q}{\partial r}$, etc. The solution to the last equation is  
\be 
q(r)=\frac{C_q}{r^3},
\label{solomega}
\ee
where $C_{q}$ is a constant to be determined from the matching between the interior and exterior solutions. 

Now we can substitute Eqs. (\ref{spher_Phi}), (\ref{spher_f}) and (\ref{spher_U}) in Eq.(\ref{con_eqn}). The resulting equation can then be expanded up to the first order in $\beta$ for the different values of $l$. In the case $l=0$, we obtain the following differential equations 
\bea
&&U_{0,rr}+ \frac{1}{r}\bigg[ 2 U_{0,r} + f_{0,r}-\Phi_{0,r} \bigg] =0,
\label{10eq0} \\ 
&&\Phi_{0,rr}+U_{0,rr}+ \frac{1}{r \left(r-2m \right)}\bigg[\left(m+r\right) \Phi_{0,r} +\left( r-m\right) \left(f_{0,r}+2U_{0,r} \right)-\frac{6}{r^4 } C^2_{q} \bigg] =0. 
\label{l0eq2} 
\eea 

In general, it is not possible to solve the above system of equations, because the number of unknown functions is greater than the number of differential equations. 
It is therefore necessary to impose an additional equation that closes the system of differential equations. Several possibilities are available. An analysis 
of the line element (\ref{lel}) shows  that in the lowest approximation of a spherically symmetric field, the metric components $g_{tt}$ and $g_{rr}$ satisfy the 
relationship $g_{rr}=-1/g_{tt}$. Consequently, we can assume the following condition $f_{0} \left(r \right)=\Phi_{0} \left(r \right)$ which allows one to easily solve the system of equations \cite{sedrakyan1968}. In addition, if at large distances, we impose the conditions $U_{0} \left(r \rightarrow \infty \right)=0$,  
$\Phi_{0} \left( r \rightarrow\infty \right)=0$ and $f_{0} \left( r \rightarrow\infty \right)=0$, we find
\bea 
&&U_{0} \left(r \right) = \frac{C_{U_{0}}}{r} ,
\label{solu0}\\
&&\Phi_{0}\left(r \right) = f_{0} \left(r \right) = \frac{ C^2_{q} +2C_{U_{0}} \ m r^2- 2C_{f_{0}} r^3 }{2 \ r^3 \left(r-2m \right)},
\label{soluf0}
\eea
where $C_{U_{0}}$ and $C_{f_{0}}$ are the integration constants of the corresponding functions.

From Eqs.(\ref{con_eqn}), we can reduce the field equations with the $l=2$ terms to:
\bea 
&&U_{2,rr}+ \frac{1}{r} \bigg[2 U_{2,r} +f_{2,r}- \Phi_{2,r}+\frac{3}{r-2 m} \left(\Phi_{2}+f _{2} \right) \bigg]=0 ,
\label{q2_U} \\
&&\Phi_{2,rr}+U_{2,rr} +\frac{r-m}{r \left(r-2m \right)} \bigg[2 U_{2,r} + f_{2,r} + \frac{1}{r-m} \left(  \left(r+m \right) \Phi_{2,r}  +3\left(f_2-\Phi_2 \right) + \frac{6 C^2_{q}}{r^4} \right) \bigg]=0 ,
\label{q2_Urr} \\
&&\Phi_{2,r}+U_{2,r}-\frac{1}{r \left(r-2m \right)} \bigg[ \left(r-3m \right) \Phi_{2} - \left(r-m \right) f_{2}  \bigg]=0 .
\label{q2_UP}
\eea 
To solve this system of equations,  we isolate $U_{2,r}$ from Eq.(\ref{q2_UP}), then we calculate $U_{2,rr}$ and substitute the resulting expressions in 
Eq.(\ref{q2_Urr}). This gives the relationship
\be 
f_{2}\left(r \right)=  \Phi_{2}\left(r \right)- \frac{3 \ C^2_{q}}{r^4}\ .
\label{f2_Phi2}
\ee 
Subsequently, the solutions of Eqs.(\ref{q2_U}), (\ref{q2_Urr}) and (\ref{q2_UP}) can be expressed as
\bea
&&\Phi_2 \left(r \right)=\frac{C^2_{q}}{2}\left(\frac{1}{m r^3}+\frac{1}{r^4}\right) - \frac{3  C_{\Phi_2}}{4} 
\left(1-\frac{2m}{r} \right)r^2 \ln \left(1- \frac{2 m}{r} \right)-\frac{ \left(3r^2-6mr-2m^2 \right) \left( r-m\right) m C_{\Phi_2}}{2 r\left(r- 2m\right)}
\label{phi2}, \\
&&f_2 \left(r \right)= \frac{C^2_{q}}{2}\left(\frac{1}{m r^3}-\frac{5}{r^4}\right)- \frac{3  C_{\Phi_2}}{4} \left(1-\frac{2m}{r} \right)r^2 \ln \left(1- \frac{2m}{r} \right)-\frac{ \left(3 r^2-6 m r-2 m^2\right)\left( r-m\right) m C_{\Phi_2}}{2 r\left(r- 2m\right)}
\label{f2}, \\
&&U_{2} \left(r \right)=-\frac{C^2_{q}}{2}\left(\frac{1}{m r^3}+\frac{2}{r^4}\right)+ \frac{3 C_{\Phi_2}}{4}  \left(1- \frac{2m^2}{r^2} \right) r^2 \ln \left(1- \frac{2m}{r} \right)+ \frac{\left(3 r^2+3 m r-2 m^2\right)m C_{\Phi_2}}{2r}
\label{U2}.
\eea

Note that due to the asymptotic flatness condition $U_{2} \left( r\rightarrow\infty \right)\rightarrow0$ the integration constant of (\ref{U2}) is related to 
$C_{\Phi_2}$ as
\be
C_{\Phi_2}=-\frac{C_{U_{2}}}{3m^2}. 
\ee
Finally,  we can rewrite the metric tensor components of the line element (\ref{lel}) as: 
\bea 
g_{00}
&=&\omega^2 e^\mu \sin^2 \theta - e^\nu\approx \beta \left(\frac{C_q}{r^3} \right)^2r^2\sin^2 \theta -\left( 1-\frac{2 m}{r} \right)\left[1+ \beta \langle \Phi_{0} \left(r \right)+\Phi_{2} \left(r \right) P_{2} \left(\cos \theta \right) \rangle \right] \\ \nonumber 
&=&- \left(1- \frac{2m}{r} \right)
\bigg\{1+ \beta \bigg\langle \frac{ C^2_{q} +2C_{U_{0}} \ m r^2- 2C_{f_{0}} r^3 }{2 \ r^3 \left(r-2m \right)}
\\ \nonumber
&+& \bigg[\frac{C^2_{q}}{2}\left(\frac{1}{m r^3}+\frac{1}{r^4}\right) - \frac{3  C_{\Phi_2}}{4} 
\left(1-\frac{2m}{r} \right)r^2 \ln \left(1- \frac{2 m}{r} \right) \\ \nonumber
&-& \frac{ \left(3r^2-6mr-2m^2 \right) \left( r-m\right) m C_{\Phi_2}}{2 r\left(r- 2m\right)}
\bigg]P_{2}\left(\cos \theta \right)-\left(1- \frac{2m}{r} \right)^{-1}\frac{C^2_{q}}{r^4}\sin^2\theta \bigg\rangle \bigg\}
\label{g00}, \\
g_{11}&=&e^\lambda\approx\left( 1-\frac{2 m}{r} \right)^{-1}\left[ 1- \beta \langle f_{0} \left(r \right)+f_{2} \left(r \right) P_{2} \left(\cos \theta \right)\rangle \right] \\ \nonumber
&=&\left( 1-\frac{2m}{r} \right)^{-1}
\bigg\{1-\beta \bigg\langle \frac{ C^2_{q} +2C_{U_{0}} \ m r^2- 2C_{f_{0}} r^3 }{2 \ r^3 \left(r-2m \right)}\\ \nonumber
&-&\bigg[\frac{C^2_{q}}{2}\left(\frac{1}{m r^3}-\frac{5}{r^4}\right)- \frac{3  C_{\Phi_2}}{4} \left(1-\frac{2m}{r} \right)r^2 \ln \left(1- \frac{2m}{r} \right)\\ \nonumber
&-&\frac{ \left(3 r^2-6 m r-2 m^2\right)\left( r-m\right) m C_{\Phi_2}}{2 r\left(r- 2m\right)}\bigg] P_{2}\left(\cos \theta \right) \bigg\rangle \bigg\}, \\
g_{22}&=&e^\mu\approx r^2\left[1+\beta \langle U_{0} \left(r \right)+U_{2} \left(r \right) P_{2} \left(\cos \theta \right) \rangle \right] \\ \nonumber
&=&r^2 \bigg\{1+\beta \bigg\langle \frac{C_{U_{0}}}{r} + \bigg[
-\frac{C^2_{q}}{2}\left(\frac{1}{m r^3}+\frac{2}{r^4}\right) + \frac{3 C_{\Phi_2}}{4}  \left(1- \frac{2m^2}{r^2} \right) r^2 \ln \left(1- \frac{2m}{r} \right)\\ \nonumber
&+& \frac{\left(3 r^2+3 m r-2 m^2\right)m C_{\Phi_2}}{2r} \bigg]P_{2}\left(\cos \theta \right) \bigg\rangle
\bigg\}, \\
g_{33}&=&g_{22}\sin^2 \theta, \\
g_{30}&=&g_{03}=\omega e^\mu \sin^2 \theta \approx\frac{C_{q}\sqrt{\beta}}{r}\sin^2 \theta.
\eea
All the constants are to be determined by matching the corresponding interior solution on the surface of the star.


\section{The relation between the Hartle-Thorne and the Sedrakyan-Chubaryan metrics}
\label{sec:tra}

In general, to establish the equivalence between two spacetimes in an invariant way it is necessary to perform a detailed analysis of the corresponding 
curvature tensors and their covariant derivatives \cite{mac15}. The problem can be simplified significantly, if it is possible to find the explicit diffeomorphism
that relates the two spacetimes. In the case that the spacetimes are approximate solutions of the field equations, the problem simplifies even further because the coordinate transformation must be valid only approximately. This is the case we are analyzing in the present work. 

In order to compare the Sedrakyan-Chubaryan solution with the Hartle-Thorne solution, we will find a coordinate transformation so that both solutions are written in the same coordinates. A close examination of the Sedrakyan-Chubaryan solution shows that it is indeed possible when one chooses the radial coordinate transformation of the type 
\be
r\rightarrow r\left(1-\frac{\beta}{2} U_0(r)\right)\ ,
\ee
and keeps the remaining coordinates unchanged. Notice that the practical effect of this transformation is to absorb the function  $ U_0(r)$. This means that, 
without loss of generality, we can set  $U_0(r)=0$ (or, equivalently, $C_{U_0}=0$) in the Sedrakyan-Chubaryan solution and thus it becomes equivalent to the
Hartle-Thorne solution, up to a redefinition of the constants entering the metric. Indeed, we now have only three integration constants, namely,
 $C_{f_0}$,  $C_{q}$ and $C_{\Phi_2}$ which are directly related to the total mass, angular momentum and quadrupole moment of the Hartle-Thorne solution. 
In fact, by comparing the $g_{tt}$ and $g_{t\phi}$ components of the metric tensor, we obtain
\be
M=m+ \frac{\beta}{2}C_{f_0},
\label{M}
\ee
\be 
J=-\frac{\sqrt{\beta}}{2}C_{q},
\label{J}
\ee
\be 
Q=\frac{\beta}{2}\bigg(\frac{C_{q}^2}{2m}-\frac{4m^5C_{\Phi_2}}{5} \bigg).
\label{Q}
\ee
Notice that $M$ in the Hartle-Thorne solution is actually composed of two terms, $M=m+\delta m$, where $m$ is the ``static mass" 
and $\delta m$ is the contribution due to the rotation 
of the source. This means that in fact the last equations relate four constants of the
 Hartle-Thorne solution with four constants of the  Sedrakyan-Chubaryan
solution, implying that the inverse transformation is well defined. This proves the mathematical and physical equivalence of the two spacetimes up to the 
first order in the quadrupole moment $Q$ and the second order in the angular momentum $J$.

There is an additional way to prove the equivalence of two spacetimes, namely, in terms of their multipole moments. 
In fact, it has been proved that two spacetimes with the same set of
multipole moments are isomorphic to each other (see, for instance, \cite{quev90} for a review on this issue). In the case of approximate solutions, one can assure that two spacetimes are isomorphic if they have the same set of multipole moments up to the validity order of the approximation. For the approximate spacetimes under consideration in this work, this means that the  Hartle-Thorne solution and the  Sedrakyan-Chubaryan are equivalent if their mass, angular momentum and quadrupole moment are the same. 
In \cite{George} and \cite{Ryan1995}, it has been shown that in the approximate case, the moments can be derived from the explicit expression of the $g_{tt}$ metric component. For the Hartle-Thorme metric we obtain   
\be 
g^{HT}_{tt}= -1+\frac{2M}{R}+\frac{2}{3}\frac{J^2}{R^4}+ 
\left[ \frac{2\left( Q-\frac{2J^2}{M}\right)}{R^3}+\frac{2\left( QM-\frac{4J^2}{3}\right)}{R^4}\right]  P_{2}\left(\cos \Theta \right),
\label{mHTtt}
\ee
whereas the corresponding expression for the Sedrakyan-Chubaryan metric reduces to 
\be 
g^{SC}_{tt}=-1+\frac{2m+\beta C_{f_0}}{R}-\frac{\beta m C_{U_0}}{R^2}+\frac{1}{6}\frac{\beta C^{2}_{q}}{R^4}- \beta \left(
\frac{\frac{1}{2}\frac{C^2_{q}}{m}+\frac{4}{5}m^{5} C_{\Phi_2}}{R^3}+
\frac{\frac{1}{6}C^2_{q}+\frac{4}{5}m^{6} C_{\Phi_2}}{R^4}
\right) P_{2}\left(\cos \Theta \right).
\label{mSCtt}
\ee
A comparison of Eqs.(\ref{mHTtt}) and (\ref{mSCtt}) shows that if we define the constants as 
\be
M=m+ \frac{\beta}{2}C_{f_0},
\label{multiM}
\ee
\be 
C_{U_0}=0, 
\label{multiCU0}
\ee
\be 
J=-\frac{\sqrt{\beta}}{2}C_{q},
\label{multiJ}
\ee
\be 
Q=\frac{\beta}{2}\bigg(\frac{C_{q}^2}{2m}-\frac{4m^5C_{\Phi_2}}{5} \bigg)\ ,
\label{multiQ}
\ee
then the moments are exactly the same. This proves the equivalence of the two metrics up to the fourth order in $1/R$.  


\section{Conclusions}
\label{sec:con}

In this work, we reviewed the original papers by Hartle (1967) and Hartle and Thorne (1968), and discussed their main properties, extensions and modifications. 
We revisited the results of Sedrakyan and Chubaryan (1968) for the metric that describes the  exterior field of an axially symmetric mass distribution. 
Using a perturbation procedure, we derived the Sedrakyan-Chubaryan solution explicitly which includes several integration constants. Instead of 
using the interior Sedrakyan-Chubaryan solution in order to find the integration constants, we compare the exterior metric with the 
exterior Hartle-Thorne spacetime solution in the same coordinates. As a result, we obtain a set of simple algebraic expressions relating 
the main parameters of the Hartle-Thorne metric with the integration constants of the Sedrakyan-Chubaryan solution. Alternatively, 
we calculated the relevant multipole moments of both solutions, and showed that they are the same. 
In this way,
we also proved the mathematical and physical equivalence of the two spacetimes. 

We conclude that the Sedrakyan-Chubaryan solution can be considered as an alternative approach to describe the gravitational field of 
a slightly deformed stationary axially symmetric mass distribution in the slow rotation approximation. 
Moreover, the Sedrakyan-Chubaryan solution with its internal counterpart can be applied to various astrophysical problems together 
with the Hartle-Thorne solution on equal rights. 

On the other hand, in a previous work \cite{bqr12} it was shown that the Hartle-Thorne formalism for the approximate 
description of rotating mass distributions is equivalent to the Fock-Abdildin approach. The last one, however,
allows us to interpret the parameters of the interior solution in terms of physical quantities like the rotational kinetic 
energy or the mutual gravitational attraction between the particles of the source. Therefore, the results obtained in this work 
imply that it should be possible to find a direct relationship between the interior Sedrakyan-Chubaryan solution 
and the corresponding counterpart in the Fock-Abdildin approach. 

It is interesting that different approaches that were developed independently in different places and under diverse circumstances 
turn out to be equivalent from a mathematical point of view. It would interesting to perform a more detailed analysis of 
all the physical characteristics of each approach in order to propose a unique formalism that would incorporate the advantages 
of all the known approaches. 

All analytical computations in this work have been performed with the help of the Mathematical Package Maple 18.

\section*{Acknowledgements}

We thank an anonymous referee for interesting remarks regarding multipole moments. 
The authors acknowledge the support of the grants No. 3101/GF4 IPC-11,No.  F.0679 0073/PTsF and K.B. acknowledges the grants for the university best  teachers-2015 and talented young scientists 2015-2016 of the Ministry of Education and Science of the Republic of Kazakhstan. K.B. is grateful to Academician D.M.~Sedrakyan for providing a copy of his doctor dissertation. This work was partially supported by DGAPA-UNAM, Grant No. 113514, and Conacyt, Grant No. 166391.


\end{document}